\documentstyle[12pt,prd,aps,epsfig,preprint]{revtex}
\tightenlines
\begin{document}
\date{\today}

\title{$\sigma$-MESON AND $\omega-\rho$ MIXING EFFECTS IN $\omega\rightarrow \pi^+\pi^-\gamma$ DECAY}

\author{A. Gokalp~\thanks{agokalp@metu.edu.tr},  A. Kucukarslan, S. Solmaz and O. Yilmaz~\thanks{oyilmaz@metu.edu.tr}
\address{ Physics Department, Middle East Technical University,
06531 Ankara, Turkey}} \maketitle

\begin{abstract}
We calculate the branching ratio of $\omega\rightarrow
\pi^+\pi^-\gamma$ decay in a phenomenological framework in which
the contributions of VMD, chiral loops, $\sigma$-meson
intermediate state amplitudes and the effects of $\omega-\rho$
mixing are considered. We conclude that the $\sigma$-meson
intermediate state amplitude and $\omega-\rho$ mixing make
substantial contribution to the branching ratio.
\end{abstract}

~~~~~~~~
\vspace{0.5cm}

The radiative decays of neutral vector mesons into a single photon
and a pair of neutral pseudoscalar mesons have been a subject of
continuous interest both theoretically and experimentally since
their study provides tests for the theoretical ideas about the
interesting mechanisms and new physics features involved in these
decays. The value
$B(\omega\rightarrow\pi^{0}\pi^{0}\gamma)=(6.6^{+1.4}_{-0.8}\pm
0.6)\times 10^{-5}$ for the branching ratio of the
$\omega\rightarrow \pi^0\pi^0\gamma$ decay  obtained  in the
recent experimental study of $\rho\rightarrow\pi^{0}\pi^{0}\gamma$
and $\omega\rightarrow\pi^{0}\pi^{0}\gamma$ decays by SND
Collaboration \cite{R1} was larger than the theoretical estimates
of this branching ratio to date which therefore requires
reexamination of the mechanism of this decay.

The theoretical study of $\omega\rightarrow \pi\pi\gamma$ decays
was initiated by Singer \cite{R2} who postulated the sequential
vector meson decay mechanism $\omega\rightarrow
(\rho)\pi\rightarrow\pi\pi\gamma$ involving the dominance of the
intermediate vector meson contribution (VMD). Bramon et al.
\cite{R3} also considered the contribution of intermediate vector
mesons to the vector meson decays into two pseudoscalars and a
single photon $V\rightarrow PP'\gamma$ using standard Lagrangians
obeying the SU(3)-symmetry. Bramon et al. \cite{R4} later studied
various such decays within the framework of chiral effective
Lagrangians using chiral perturbation theory and they shown that
there is no tree-level contribution to the amplitudes of such
decay processes and the one-loop contributions are finite. Guetta
and Singer \cite{R5} in a recent work combined all the
improvements on the simple Born term of VMD mechanism. They
considered $\omega-\rho$ mixing, momentum dependence of
intermediate state $\rho$-meson width, the inclusion of the chiral
loop amplitude as given by Bramon et al. \cite{R4}, and using the
resulting  full amplitude which includes all these effects they
obtained the theoretical result
$B(\omega\rightarrow\pi^{0}\pi^{0}\gamma)=(4.6\pm 1.1)\times
10^{-5}$ for the branching ratio of the
$\omega\rightarrow\pi^{0}\pi^{0}\gamma$ decay which is seriously
less than the latest experimental result \cite{R1}.

Another interesting contribution to the decay mechanism of this
decay may involve $\sigma$-meson as an intermediate state. The
existence of $\sigma$-meson has long been controvertial, however,
recently a large number of analyzes point to its existance
\cite{R6}. The recent Fermilab E791 experiment found strong direct
experimental evidence for $\sigma$-meson in the measured
$D^+\rightarrow\sigma\pi^+\rightarrow \pi\pi\pi$ decay \cite{R7}.
Therefore this important meson must be included in the analyzes of
hadronic processes and its contribution to the mechanisms of such
processes should be examined. However, the nature and the  quark
substructure of $\sigma$-meson have not been established yet,
whether it is a conventional  $q\bar{q}$ state or a $\pi\pi$
resonance has been a subject of debate. Therefore, analysis of the
role of $\sigma$-meson in $V\rightarrow PP\gamma$ decays may also
provide information about the controvertial nature of
$\sigma$-meson.

The three of the present authors in an attempt to explain the
latest experimental result of the branching ratio of
$\omega\rightarrow\pi^{0}\pi^{0}\gamma$ decay, reconsidered the
decay mechanism of this decay in a phenomenological framework in
which the contributions of VMD and chiral loop amplitudes, the
effects of the $\omega-\rho$ mixing, and the contribution of
$\sigma$-meson intermediate state amplitude are included, and as
the result of their analysis they concluded that $\sigma$-meson
intermediate state should be included in the decay mechanism of
$\omega\rightarrow\pi^{0}\pi^{0}\gamma$ decay  in order to explain
the latest experimental result, and utilizing the experimental
value of the branching ratio of
$\omega\rightarrow\pi^{0}\pi^{0}\gamma$ decay, which resulted in a
quadric equation for g$_{\omega\sigma\gamma}$, estimated the
coupling constant g$_{\omega\sigma\gamma}$ as
g$_{\omega\sigma\gamma}=0.11$ and g$_{\omega\sigma\gamma}=-0.21$
\cite{R8}. An essential assumption of that work was that there is
no SU(3) vector meson-sigma-gamma vertex, therefore
$\omega\sigma\gamma$-vertex cannot be related to the
$\rho\sigma\gamma$-vertex. The effects of $\sigma$-meson in the
mechanism of radiative $\rho^0$-meson decays is included by
assuming that $\sigma$-meson couples to $\rho^0$-meson through the
pion-loop \cite{R8,R9}. However, in the mechanism of
$\omega\rightarrow\pi^{0}\pi^{0}\gamma$ decay a
$\omega\sigma\gamma$-vertex was assumed which may be considered as
representing the effective final state interactions in
$\pi\pi$-channel \cite{R8}. The small value of the coupling
constant $g_{\omega\sigma\gamma}$ obtained resulted in a change in
the Born amplitude of $\omega\rightarrow\pi^{0}\pi^{0}\gamma$
decay which was of the same order of magnitude as it is typical of
final state interactions. Indeed, Levy and Singer \cite{R10} in a
study of the $\omega\rightarrow\pi^{0}\pi^{0}\gamma$ decay using
dispersion-theoretical approach shown that final state
interactions resulting in a decay rate of the same order of
magnitude as the one calculated from the Born term can be
parametrized with the effective-pole approximation.

In order to investigate  the proposed role of $\sigma$-meson in
radiative $\rho^0$- and $\omega$- meson decays further and to
obtain more results that can be tested by experiment, we study in
this work $\omega\rightarrow\pi^{+}\pi^{-}\gamma$ decay in a
phenomenological approach by considering the vector meson
dominance, chiral loops, and $\sigma$-meson intermediate state
amplitudes as well as the effects of $\omega-\rho$ mixing. We
calculate the branching ratio of this decay and we obtain the
photon spectra for the branching ratio of
$\omega\rightarrow\pi^{+}\pi^{-}\gamma$ decay which can be tested
experimentally. If only Born term VMD amplitude is used one has
the relation $\Gamma(\omega\rightarrow\pi^0\pi^0\gamma)=
(1/2)\Gamma(\omega\rightarrow\pi^{+}\pi^{-}\gamma)$ which follows
from charge conjugation invariance to order $\alpha$ which imposes
pion pairs of even angular momentum as shown by Singer \cite{R2}.
Since the factor (1/2) holds to the first order in $\alpha$, the
present calculation is also of interest since the amplitude
resulting from the assumed decay mechanism for $\omega$- and
$\rho$-meson decays in the present work contains terms of order
$e^3$.

Our calculation is based on the Feynman diagrams shown in Fig. 1
for $\omega\rightarrow\pi^+\pi^-\gamma$ decay and in Fig. 2 for
$\rho^{0}\rightarrow\pi^{+}\pi^{-}\gamma$ decay. The direct terms
shown in the diagrams in Fig. 2 a, b, c are required to establish
the gauge invariance. The $\omega\rho\pi$-vertex is described by
the Wess-Zumino anomaly term of the chiral Lagrangian \cite{R11}
\begin{eqnarray} \label{e1}
{\cal L}^{eff}_{\omega\rho\pi}=g_{\omega\rho\pi}
\epsilon^{\mu\nu\alpha\beta}\partial_{\mu}\omega_{\nu}
\partial_{\alpha}\vec{\rho}_{\beta}\cdot\vec{\pi}~~.
\end{eqnarray}
This effective Lagrangian also defines the coupling constant
$g_{\omega\rho\pi}$ which  was determined by Achasov et al.
\cite{R12} through an experimental analysis as
$g_{\omega\rho\pi}=(14.4\pm 0.2)~~GeV^{-1}$.
 Similarly we describe
the $\rho\pi\gamma$-vertex with the effective Lagrangian
\begin{eqnarray}\label{e2}
{\cal L}^{eff}_{\rho\pi\gamma}=g_{\rho\pi\gamma}
\epsilon^{\mu\nu\alpha\beta}\partial_{\mu}\vec{\rho}_{\nu}\cdot
\vec{\pi}\partial_{\alpha}A_{\beta}~~,
\end{eqnarray}
and we use  the experimental partial width  of the radiative
$\rho\rightarrow\pi\gamma$ decay \cite{R13} to deduce the coupling
constant $g_{\omega\pi\gamma}$  as $g_{\rho\pi\gamma}=(0.274\pm
0.035)~~GeV^{-1}$. The $\omega\sigma\gamma$-vertex is described by
the effective Lagrangian
\begin{eqnarray}\label{e3}
{\cal
L}^{eff}_{\omega\sigma\gamma}=\frac{e}{M_\omega}g_{\omega\sigma\gamma}
\partial^\alpha\omega^\beta(\partial_{\alpha}A_\beta-\partial_{\beta}A_\alpha)\sigma~~.
\end{eqnarray}
The coupling constant $g_{\omega\sigma\gamma}$ was estimate as
$g_{\omega\sigma\gamma}=0.11$ and $g_{\omega\sigma\gamma}=-0.21$
by three of the present authors \cite{R8} in their analysis of
$\omega\rightarrow\pi^0\pi^0\gamma$ decay in a similar
phenomenological framework. We describe the  $\sigma\pi\pi$-vertex
by the effective Lagrangian
\begin{eqnarray} \label{e4}
{\cal L}^{eff}_{\sigma\pi\pi}=
\frac{1}{2}g_{\sigma\pi\pi}M_{\sigma}\vec{\pi}\cdot\vec{\pi}\sigma~~,
\end{eqnarray}
and using the experimental values for $M_{\sigma}$ and
$\Gamma_\sigma$ as  $M_{\sigma}=(483\pm 31)$ MeV and
$\Gamma_\sigma=(338\pm 48)$ MeV \cite{R7} where statistical and
systematic errors are added in quadrature we determine the strong
coupling constant $g_{\sigma\pi\pi}$ as $g_{\sigma\pi\pi}=(5.3\pm
0.55)$. The $\rho\pi\pi$-vertex is described by the effective
Lagrangian
\begin{eqnarray}\label{e5}
{\cal L}^{eff}_{\rho\pi\pi}=g_{\rho\pi\pi}
\vec{\rho}_{\mu}\cdot(\partial^{\mu}\vec{\pi}\times\vec{\pi})~~,
\end{eqnarray}
and the experimental decay width of the decay
$\rho\rightarrow\pi\pi$ \cite{R13} yields the value
$g_{\rho\pi\pi}=(6.03\pm 0.02)$ for the coupling constant
$g_{\rho\pi\pi}$.The effective Lagrangians ${\cal
L}^{eff}_{\sigma\pi\pi}$ and ${\cal L}^{eff}_{\rho\pi\pi}$ result
from an extension of the $\sigma$ model where the isovector $\rho$
is included through a Yang-Mills local gauge theory based on
isospin with the vector meson mass generated through the Higgs
mechanism \cite{R14}.

Meson-meson interactions were studied by Oller and Oset \cite{R15}
using the standard chiral Lagrangian in lowest order of chiral
perturbation theory.We use their result for the isospin I=0
$\pi^+\pi^-\rightarrow\pi^+\pi^-$ amplitude that we need in the
loop diagrams in Fig. 2 b, thus we neglect the small I=2
amplitude. We note that as shown by Oller \cite{R16} due to gauge
invariance the off-shell parts of the amplitudes, that should be
kept inside the loop integration, do not contribute, and
consequently the amplitude ${\cal
M}_\chi(\pi^+\pi^-\rightarrow\pi^+\pi^-)$ factorizes in the
expression for the loop diagrams.

In our calculation of the invariant amplitude, we make the
replacement $q^2-M^2\rightarrow q^2-M^2+iM\Gamma$ in $\rho$-meson
and $\sigma$-meson propagators. We use for $\sigma$-meson the
momentum dependent width that follows from Eq. 4
\begin{eqnarray}\label{e6}
\Gamma_\sigma (q^2)=\Gamma_\sigma
\frac{M_\sigma^2}{q^2}\sqrt{\frac{q^2-4M_\pi^2}{M_\sigma^2-4M_\pi^2}}
\theta(q^2-4M_\pi^2)~~,
\end{eqnarray}
and for $\rho$-meson we use the following momentum dependent width
as conventionally adopted \cite{R17}
\begin{eqnarray}\label{e7}
\Gamma_\rho (q^2)=\Gamma_\rho
\frac{M_\rho}{\sqrt{q^2}}\left(\frac{q^2-4M_\pi^2}{M_\rho^2-4M_\pi^2}\right)^{3/2}
\theta(q^2-4M_\pi^2)~~.
\end{eqnarray}

Loop integrals similar to the ones appearing in Figs. 1 and 2 were
evaluated by Lucio and Pestiau \cite{R18} using dimensional
regularization. We use their results and, for example, we express
the contribution of the pion-loop amplitude corresponding to
$\rho^0\rightarrow(\pi^+\pi^-)\gamma\rightarrow\pi^+\pi^-\gamma$
reaction in Fig. 2 b as
\begin{eqnarray}\label{e8}
{\cal M_\pi}=-\frac{e~g_{\rho\pi\pi}{\cal
M}_{\chi}(\pi^+\pi^-\rightarrow\pi^+\pi^-)}{2\pi^{2}M_{\pi}^{2}}I(a,b)
\left[(p\cdot k)(\epsilon\cdot u)-(p\cdot\epsilon)(k\cdot
u)\right]~~,
\end{eqnarray}
where $a=M_\rho^2/M_\pi^2$, $b=(p-k)^2/M_\pi^{2}$, ${\cal
M}_{\chi}=-(2/f_\pi^2)(s+M_\pi^2/6)$,~ $s=(p-k)^2$,~ $f_\pi=92.4$
MeV, $p(u)$ and $k(\epsilon)$ being the momentum (polarization
vector) of $\rho$-meson and photon, respectively. A similar
amplitude corresponding to
$\rho^0\rightarrow(\pi^+\pi^-)\gamma\sigma\rightarrow\pi^+\pi^-\gamma$
reaction can also be written. The function I(a,b) is given as
\begin{eqnarray}\label{e9}
I(a,b)=\frac{1}{2(a-b)} -\frac{2}{(a-b)^{2}}\left [
f\left(\frac{1}{b}\right)-f\left(\frac{1}{a}\right)\right ]
+\frac{a}{(a-b)^{2}}\left [
g\left(\frac{1}{b}\right)-g\left(\frac{1}{a}\right)\right ]
\end{eqnarray}
where
\begin{eqnarray}\label{e10}
&&f(x)=\left \{
\begin{array}{rr}
           -\left [ \arcsin (\frac{1}{2\sqrt{x}})\right ]^{2}~,& ~~x>\frac{1}{4} \\
\frac{1}{4}\left [ \ln (\frac{\eta_{+}}{\eta_{-}})-i\pi\right
]^{2}~, & ~~x<\frac{1}{4}
            \end{array} \right.
\nonumber \\ && \nonumber \\ &&g(x)=\left \{ \begin{array}{rr}
        (4x-1)^{\frac{1}{2}} \arcsin(\frac{1}{2\sqrt{x}})~, & ~~ x>\frac{1}{4} \\
 \frac{1}{2}(1-4x)^{\frac{1}{2}}\left [\ln (\frac{\eta_{+}}{\eta_{-}})-i\pi \right ]~, & ~~ x<\frac{1}{4}
            \end{array} \right.
\nonumber \\ && \nonumber \\ &&\eta_{\pm}=\frac{1}{2x}\left [
1\pm(1-4x)^{\frac{1}{2}}\right ] ~.
\end{eqnarray}

We describe the $\omega-\rho$ mixing by an effective Lagrangian of
the form
\begin{eqnarray}\label{e11}
{\cal L}^{eff}_{\rho-\omega}=\Pi_{\rho\omega}\omega_\mu\rho^\mu~~,
\end{eqnarray}
where $\omega_\mu$ and $\rho_\mu$ denote pure isospin field
combinations. The mixing allows the transition $\omega\rightarrow
\rho$ in the process $\omega\rightarrow\pi^{+}\pi^{-}\gamma$, thus
the amplitude of the decay can be written as ${\cal M}={\cal
M}_0+\epsilon {\cal M}^\prime$ where ${\cal M}_0$ includes the
contributions coming from the diagrams shown in Fig. 1 for
$\omega\rightarrow\pi^{+}\pi^{-}\gamma$ and ${\cal M^\prime}$
represents the contributions of the diagrams in Fig. 2 for
$\rho^0\rightarrow\pi^{+}\pi^{-}\gamma$. The mixing parameter
$\epsilon$ is given as \cite{R17}
\begin{eqnarray}\label{e12}
\epsilon=\frac{\Pi_{\rho\omega}}{M_\omega^2-M_\rho^2
+iM_\rho\Gamma_\rho-iM_\omega\Gamma_\omega}~~.
\end{eqnarray}
O`Connell et al. \cite{R17} determined $\Pi_{\rho\omega}$ from
fits to $e^+e^-\rightarrow \pi^+\pi^-$ data as
$\Pi_{\rho\omega}=(-3800\pm 370)~~ MeV^2$ from which the mixing
parameter is obtained as $\epsilon=(-0.006+i 0.036)$. Another
effect of $\omega-\rho$ mixing is that it modifies the
$\rho$-propagator in diagrams in Fig. 1 a
\begin{eqnarray}\label{e13}
\frac{1}{D_\rho(s)}\rightarrow
\frac{1}{D_\rho(s)}\left(1+\frac{g_{\omega\pi\gamma}}{g_{\rho\pi\gamma}}
\frac{\Pi_{\rho\omega}}{D_\rho(s)}\right)
\end{eqnarray}
where $D_\rho(s)=s-M_\rho^2+iM_\rho\Gamma_\rho(s)$. This effect is
relevant since according to SU(3) relation
$g_{\omega\pi\gamma}/g_{\rho\pi\gamma}=3$ it makes a sizable
contribution.

We calculate the invariant amplitude ${\cal M}$(E$_{\gamma}$,
E$_{1}$) this way for the decay
$\omega\rightarrow\pi^{+}\pi^{-}\gamma$ from the corresponding
Feynman diagrams shown in Fig. 1 and 2 for the decays
$\omega\rightarrow\pi^+\pi^-\gamma$ and
$\rho^{0}\rightarrow\pi^{+}\pi^{-}\gamma$, respectively.The
differential decay probability for an unpolarized $\omega$-meson
at rest is then given as
\begin{eqnarray}\label{e14}
\frac{d\Gamma}{dE_{\gamma}dE_{1}}=\frac{1}{(2\pi)^{3}}~\frac{1}{8M_{\omega}}~
\mid {\cal M}\mid^{2} ,
\end{eqnarray}
where E$_{\gamma}$ and E$_{1}$ are the photon and pion energies
respectively. We perform an average over the spin states of the
vector meson and a sum over the polarization states of the photon.
The decay width is then obtained by integration
\begin{eqnarray}\label{e15}
\Gamma=\int_{E_{\gamma,min.}}^{E_{\gamma,max.}}dE_{\gamma}
       \int_{E_{1,min.}}^{E_{1,max.}}dE_{1}\frac{d\Gamma}{dE_{\gamma}dE_{1}}~~.
\end{eqnarray}
Although the minimum photon energy is E$_{\gamma, min.}=0$, in our
calculations it is taken as E$_{\gamma, min.}=30$ MeV because of
the presence of bremsstrahlung amplitude, and the maximum photon
energy is given as
$E_{\gamma,max.}=(M_{\omega}^{2}-4M_{\pi}^{2})/2M_{\omega}$=341
MeV. The maximum and minimum values for pion energy E$_{1}$ are
given by
\begin{eqnarray}\label{e16}
\frac{1}{2(2E_{\gamma}M_{\omega}-M_{\omega}^{2})} [
-2E_{\gamma}^{2}M_{\omega}+3E_{\gamma}M_{\omega}^{2}-M_{\omega}^{3}
 ~~~~~~~~~~~~~~~~~~~~~~~~~~~~ \nonumber \\
\pm  E_{\gamma}\sqrt{(-2E_{\gamma}M_{\omega}+M_{\omega}^{2})
       (-2E_{\gamma}M_{\omega}+M_{\omega}^{2}-4M_{\pi}^{2})}~] ~.
\nonumber
\end{eqnarray}

As the result of our calculation, if we use the full amplitudes
resulting from the Feynman diagrams in Fig. 1 and in Fig. 2, we
obtain for the branching ratio of
$\omega\rightarrow\pi^{+}\pi^{-}\gamma$ decay the value
$B(\omega\rightarrow\pi^{+}\pi^{-}\gamma)=0.43\times 10^{-3}$
using the coupling constant $g_{\omega\sigma\gamma}=0.11$ and
$B(\omega\rightarrow\pi^{+}\pi^{-}\gamma)=0.67\times 10^{-3}$ if
we use the coupling constant $g_{\omega\sigma\gamma}=-0.21$. These
values are consistent with the present experimental upper limit
$B(\omega\rightarrow\pi^{+}\pi^{-}\gamma)<3.6\times 10^{-3}$
\cite{R13}. If we use only the VMD amplitude for
$\omega\rightarrow\pi^{+}\pi^{-}\gamma$ decay we obtain the
branching ratio $B(\omega\rightarrow\pi^{+}\pi^{-}\gamma)=
7.2\times 10^{-5}$. If we use the VMD amplitude for
$\omega\rightarrow\pi^{+}\pi^{-}\gamma$ and the bremsstrahlung
amplitude for $\rho^0\rightarrow\pi^{+}\pi^{-}\gamma$, as a result
of $\omega-\rho$ mixing we obtain the branching ratio
$B(\omega\rightarrow\pi^{+}\pi^{-}\gamma)=0.46\times 10^{-3}$. On
the other hand, If we use VMD and $\sigma$-meson intermediate
state amplitudes for $\omega\rightarrow\pi^{+}\pi^{-}\gamma$ decay
and do not consider $\omega-\rho$ mixing, the resulting branching
ratio is $B(\omega\rightarrow\pi^{+}\pi^{-}\gamma)=0.13\times
10^{-3}$ for $g_{\omega\sigma\gamma}=0.11$ and
$B(\omega\rightarrow\pi^{+}\pi^{-}\gamma)=0.12\times 10^{-3}$ for
$g_{\omega\sigma\gamma}=-0.21$. The values of the branching ratio
$B(\omega\rightarrow\pi^{+}\pi^{-}\gamma)$ resulting from the
different amplitudes for the coupling constant
$g_{\omega\sigma\gamma}=0.11$ are shown in Table 1 for comparison.
These values show the importance of not only $\omega-\rho$ mixing
but also of the $\sigma$-meson intermediate state amplitude in
$\omega\rightarrow\pi^{+}\pi^{-}\gamma$ decay.

\begin{table}
\caption{The values of the branching ratio
$B(\omega\rightarrow\pi^{+}\pi^{-}\gamma)$ resulting from the VMD
amplitude, VMD amplitude and $(\omega-\rho)$ mixing with the
bremsstrahlung amplitude, $\sigma$-meson intermediate state
amplitude, VMD and $\sigma$-meson intermediate state amplitude,
and full amplitude including VMD, $\sigma$-meson intermediate
state and $(\omega-\rho)$ mixing for the coupling constant
$g_{\omega\sigma\gamma}=0.11$}
\begin{center}
\begin{tabular}{|c|c|c|c|c|c|}\hline
&&&&&\\
Amplitude & VMD & VMD+ & $\sigma$ & VMD+$\sigma$ & VMD+$\sigma$ \\
& & $(\omega-\rho)$ & & &$+(\omega-\rho)$ \\ \hline &&&&& \\
$B(\omega\rightarrow\pi^{+}\pi^{-}\gamma)\times 10^{5}$ & 7.2
& 46 & 2.5& 13 & 43\\
&&&&& \\
 \hline
\end{tabular}
\end{center}
\end{table}

The photon spectra for the branching ratio of the decay
$\omega\rightarrow\pi^{+}\pi^{-}\gamma$, which may be tested in
future experiments, are plotted in Fig. 3 for
$g_{\omega\sigma\gamma}=0.11$ and in Fig. 4 for
$g_{\omega\sigma\gamma}=-0.21$ as a function of photon energy
E$_\gamma$, and the contributions of different amplitudes are
indicated. We note that the bremsstrahlung amplitude of
$\rho^0\rightarrow\pi^{+}\pi^{-}\gamma$ decay as the result of
$\omega-\rho$ mixing affects mostly the lower part of the photon
spectra, changing it drastically, but becomes practically
negligible toward the higher photon energy part of the spectrum.
Therefore, the model and the ideas about the role of $\omega-\rho$
mixing and the importance of the contribution of $\sigma$-meson
intermediate state in $\omega\rightarrow\pi^{+}\pi^{-}\gamma$
decay presented in this work can be tested experimentally if few
events are collected at the characteristic parts of the photon
spectra.

\newpage
\begin{figure}\vspace*{1.0cm}\hspace{0.0cm}
\epsfig{figure=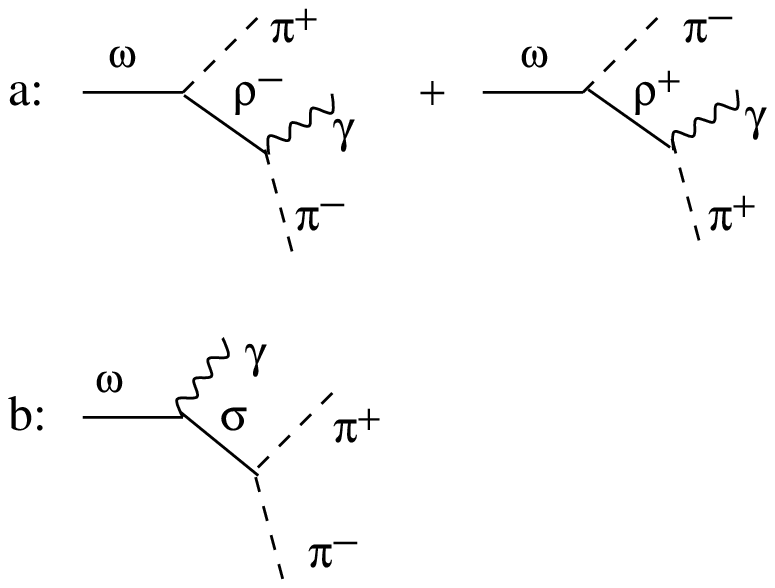,width=7cm,height=5cm,
angle=0}\vspace*{0.5cm} \caption{Feynman diagrams for the decay
$\omega\rightarrow\pi^+\pi^-\gamma$.} \label{fig1}
\end{figure}

\begin{figure}
\vspace*{1.0cm}\hspace{0.0cm}
\epsfig{figure=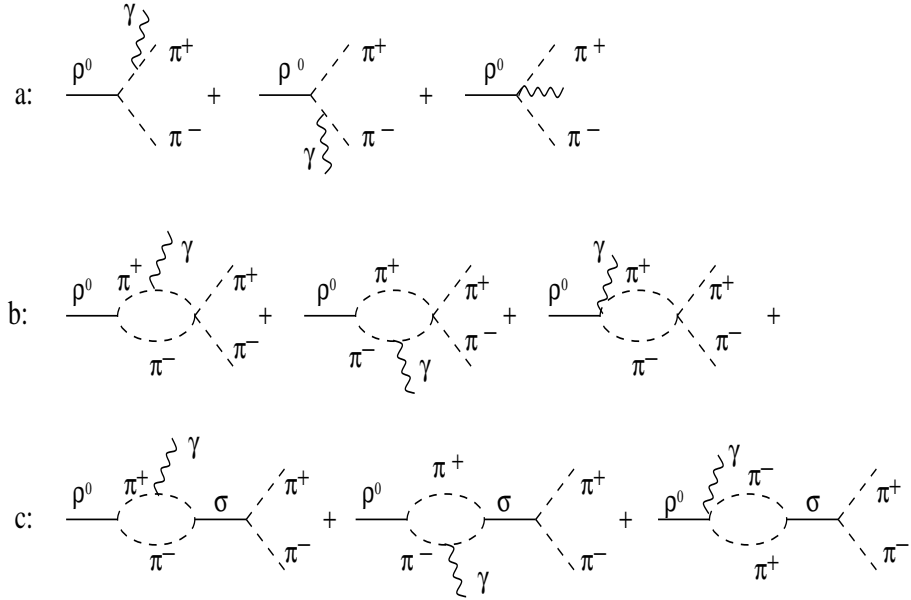,width=12cm,height=8cm,
angle=0}\vspace*{0.5cm} \caption{Feynman diagrams for the decay
$\rho^0\rightarrow\pi^+\pi^-\gamma$.} \label{fig2}
\end{figure}

\newpage
\begin{figure}\hspace{-1.5cm}
\epsfig{figure=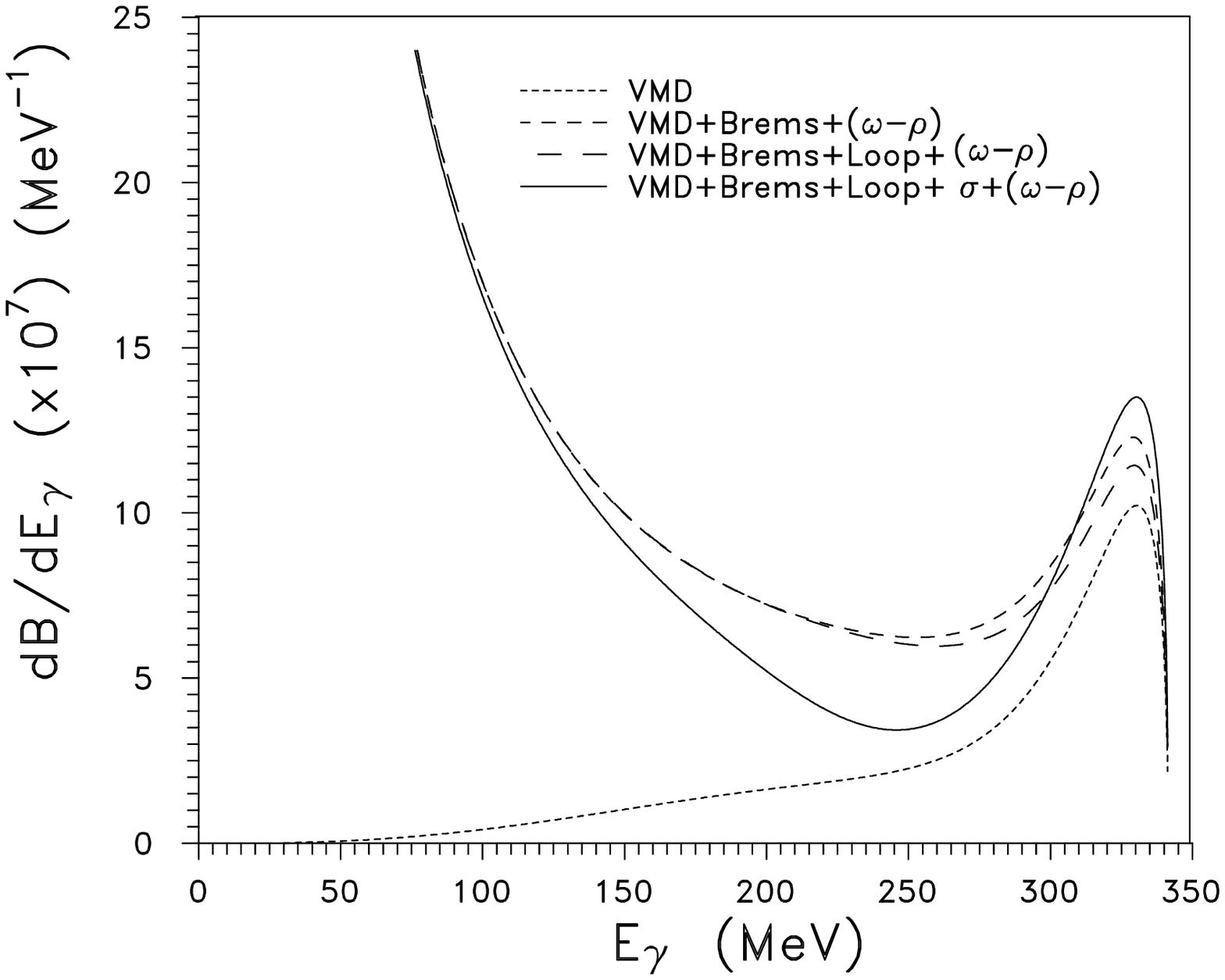,width=15cm,height=20cm} \vspace*{-4.5cm}
\caption{The photon spectra for the branching ratio of
$\omega\rightarrow\pi^+\pi^-\gamma$ decay for
$g_{\omega\sigma\gamma}=0.11$. The separate contributions
resulting from the amplitudes of VMD; VMD and bremsstrahlung with
$\omega-\rho$ mixing; VMD and bremsstrahlung , chiral loop with
$\omega-\rho$ mixing; and from the full amplitude using the
diagrams in Fig. 1 and in Fig. 2 including $\sigma$-meson
intermediate state with $\omega-\rho$ mixing.} \label{fig3}
\end{figure}

\begin{figure}\hspace{-1.5cm}
\epsfig{figure=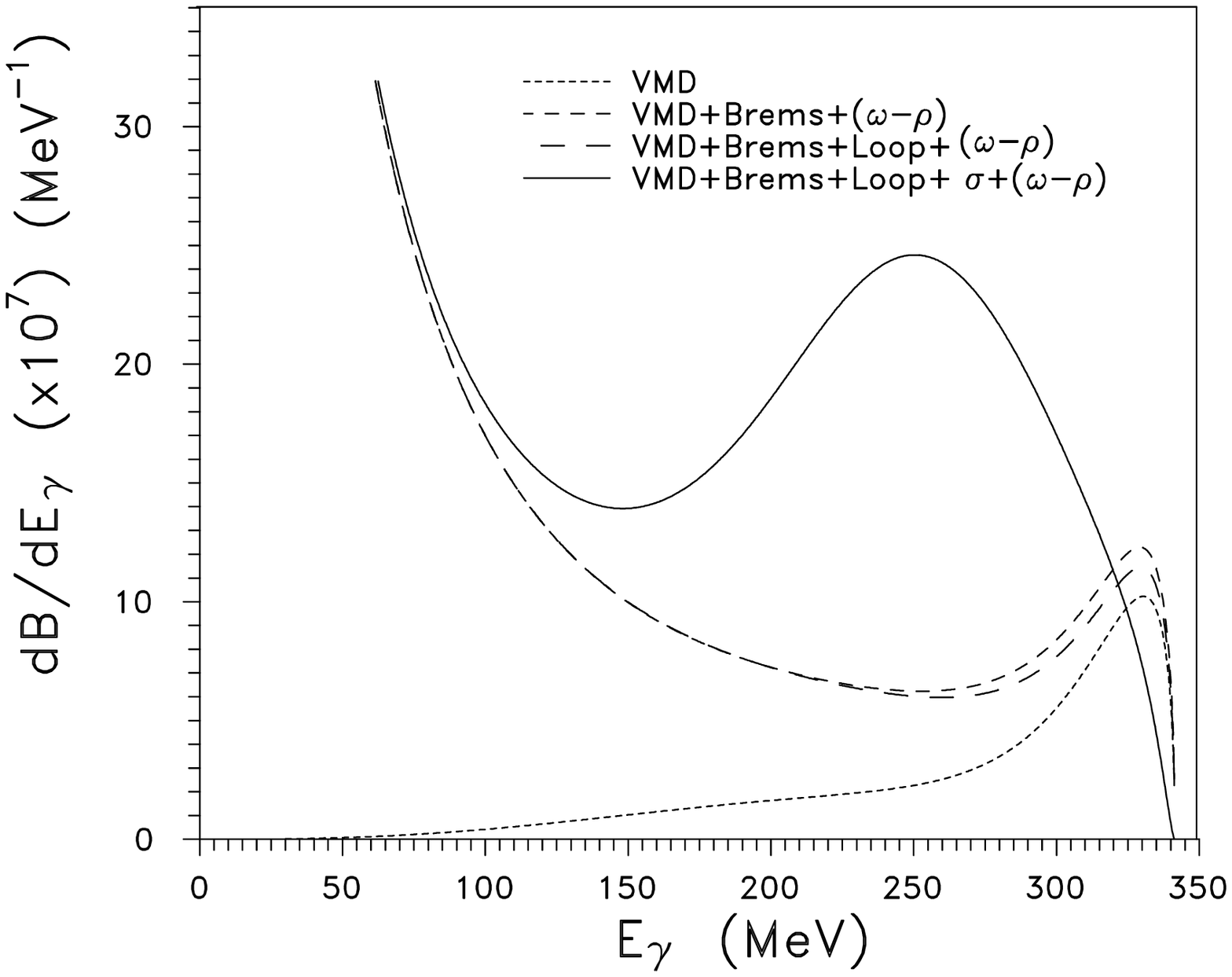,width=15cm,height=20cm} \vspace*{-4.5cm}
\caption{The photon spectra for the branching ratio of
$\omega\rightarrow\pi^+\pi^-\gamma$ decay for
$g_{\omega\sigma\gamma}=-0.21$. The separate contributions
resulting from the amplitudes of VMD; VMD and bremsstrahlung with
$\omega-\rho$ mixing; VMD and bremsstrahlung , chiral loop with
$\omega-\rho$ mixing; and from the full amplitude using the
diagrams in Fig. 1 and in Fig. 2 including $\sigma$-meson
intermediate state with $\omega-\rho$ mixing.} \label{fig4}
\end{figure}

\end{document}